\documentclass{LT23auth}
\usepackage{graphicx}

\begin{document}

\begin{frontmatter}

% Use lower case letters in the title.
\title{Diffusion of an Inhomogeneous Vortex Tangle}

\author[address1]{Makoto Tsubota\thanksref{thank1}},
\author[address1]{Tsunehiko Araki},
\author[address2]{W.F.Vinen}

\address[address1]{Department of Physics, Osaka City University,
Osaka 558-8585, Japan}
\address[address2]{School of Physics and Astronomy,
University of Birmingham, Birmingham B15 2TT, UK}

% The corresponding author should be distinguished and his email
% address and/or fax number must be given. His mailing address has to
% be complete: the proofs are send to this address around
% January 1, 2003. The address for sending proofs has to be indicated
% as "present address", if it is different from the address above.
\thanks[thank1]{ E-mail:tsubota@sci.osaka-cu.ac.jp}

\begin{abstract}
The spatial diffusion of an inhomogeneous vortex tangle is studied
numerically with the vortex filament model.
A localized initial  tangle is prepared by applying a counterflow,
and the tangle is allowed to diffuse freely after the counterflow
is turned off. Comparison with the solution of a generalization of
the Vinen equation that takes diffusion into account leads to a
very small diffusion constant, as expected from simple theoretical
considerations.  The relevance of this result to recent experiments
on the generation and decay of superfluid turbulence at very low
temperatures is discussed.
\end{abstract}

%
% write here 3 or 4 keywords separated by semicolons
%
\end{frontmatter}

\def\vx{\mbox{\boldmath $x$}}

\section{Introduction}

Recently there has been a growing interest in a comparison between
classical turbulence and quantum turbulence in
superfluid$^4$He \cite{Vinen}.
The ideal comparison requires experiments on  homogeneous quantum
turbulence  produced by a grid  towed at a steady velocity through
the helium \cite{Stalp}, and there is particular interest in the
case of very low temperatures, when there is practically no normal fluid.
Steady towing of a grid at these low temperatures  leads to severe
experimental problems, so the only existing relevant experiment used
an oscillating grid instead of a towed grid \cite{Davis}.
Any interpretation of this experiment requires an understanding
of the behaviour of inhomogeneous quantum turbulence, and this paper
aims to provide relevant evidence, based on a computer simulation
of the diffusion of a localized random vortex tangle.
The results show that diffusion in this case is very slow, as is
to be expected from simple dimensional considerations.
Low-temperature grid turbulence may involve motion on scales much
larger than that associated with a random tangle, which is of order
the vortex-line spacing, in which case diffusion would be faster.

\begin{figure}[tbhp]
\begin{minipage}{1.0\linewidth}
\begin{center}
\includegraphics[width=0.8\linewidth]{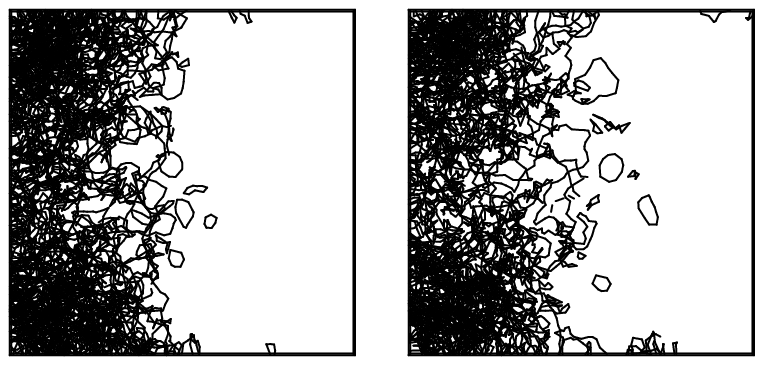}\\
 (a) \hspace{2cm} (b)\\
\end{center}
\end{minipage}

\begin{minipage}{1.0\linewidth}
\begin{center}
\includegraphics[width=0.8\linewidth]{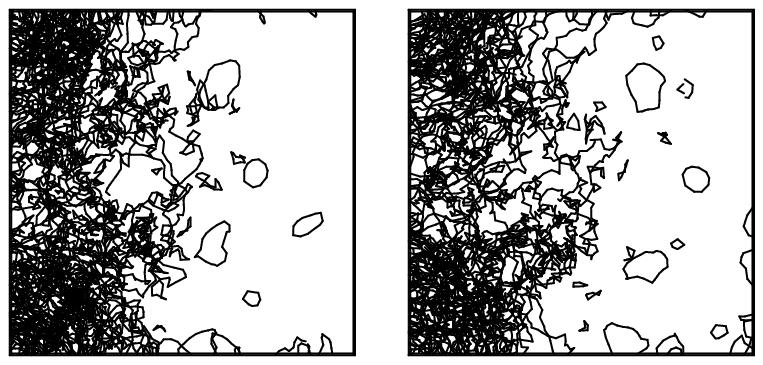}\\
 (c) \hspace{2cm} (d)\\
\end{center}
\end{minipage}
\caption{Diffusion of a vortex tangle at $t$=0 sec(a), $t$=10.0 sec(b),
$t$=20.0 sec(c) and $t$=30.0 sec(d).}
 \label{eps1}
\end{figure}

\section{Numerical calculation}
The required equation of motion of the vortices and the numerical
procedures  are described in detail in a previous paper \cite{Tsubota}.
The tangle is confined to a 1cm cube.
An initial configuration of six vortex rings is allowed to evolve
under a counterflow at 1.6K, using the local induction approximation.
Periodic boundary conditions are applied at the faces normal to the flow;
the other faces are taken as solid. When a random tangle has developed,
the counterflow is turned off and the temperature reduced to zero,
vortices with parts in the right-hand half of the cube are removed,
and the evolution of the remaining vortices is followed, now with a
fully non-local Biot-Savart dynamics. The evolution involves both
decay and diffusion.  The results are shown in Figs. 1 and 2.

\section{A model equation to describe decay and diffusion of a vortex tangle}
The decay of a homogeneous vortex tangle is described by the
Vinen's equation
\begin{equation}
dL/dt=-\chi_2\frac{\kappa}{2\pi}L^2,
\end{equation}
where $L$ is the vortex line density and $\kappa$ is the quantum of
circulation ($h/m_{4}$) \cite{Vinen57}.
The parameter $\chi_2$ depends on temperature, being about 0.3
at zeto temperature \cite{Tsubota}.
A simple generalization of this equation for an inhomogeneous system
can be obtained by adding a diffusion term.  The resulting equation
is unlikely to be rigorously correct, but it will serve for a
preliminary analysis of the experiments results.
That is,  we assume that
\begin{equation}
\frac{dL(\vx,t)}{dt} = -\chi_2 \frac{\kappa}{2\pi} {L(\vx,t)}^2
+ D \nabla^2 L(\vx, t).
\label{Vinen}
\end{equation}
where $D$ is a diffusion coefficient,  assumed constant for a given
temperature.
We see from the results in figure 2 that as time proceeds the line density
falls
near the wall at $x=0$.  This feature  is not described by Eq. (2), and it is
due presumably to an enhanced vortex decay rate in the neighbourhood of the
wall.
We take it into account by allowing $\chi_{2}$ to increase close to the wall.
As explained in Ref. \cite{Tsubota}, the simulations are based on a finite
spatial resolution, $\Delta x$.
The parameter $\chi_{2}$ is therefore increased over the range from $x=0$
to $x=\Delta x$ from the value $\chi_{2}$  to a value $\chi_{2, B}$.

We have solved Eq. (2) numerically for the situation obtaining in the
experiments.
Taking $\chi_{2}=0.3$,  and regarding $D$  and $\chi_{2,B}$ as adjustable
parameters,
we have carried out least square fits to the experimental data.
We find that at all times $D/\kappa = 0.1\pm 0.05$ and $\chi_{2,D}= 0.7\pm
0.05$.

An theoretical estimate of the likely value of $D$ can be obtained as
follows.
We assume that the tangle is random, so that there is no significant
superfluid
motion on a scale larger than the line spacing $\ell=L^{-1/2}$.
Then the only length scale relevant to the vortex motion is $\ell$ and
the only velocity scale is $\kappa/ \ell$ (we ignore logarithmic corrections
arising from line curvature). It follows that $D$ must be of order $\kappa$,
as is observed.

\begin{figure}[tbhp]
\begin{center}
\includegraphics[width=0.9\linewidth]{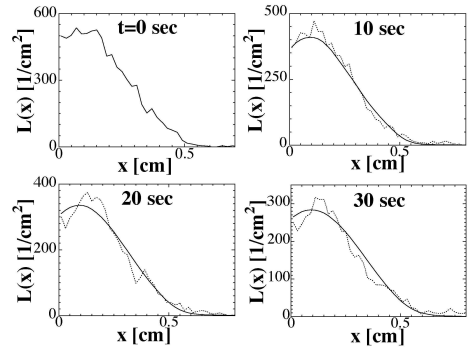}\\
\end{center}
\caption{Development of the vortex line density distribution.
The dotted line shows the results of the numerical simulation of
Fig. 1, while the solid lines are discussed in section 3.}
\label{Comparison}
\end{figure}

\section{Conclusion}

The diffusion of an inhomogeneous random vortex tangle has been studied
numerically
with the vortex filament model.
The diffusion can be described by a generalization of the Vinen equation,
with a diffusion constant close to  $0.1 \kappa$.

This result may be relevant to the generation of superfluid turbulence
with an oscillating grid \cite{Davis}. If the tangle that is produced has
associated
with it no motion on a length scale larger than the line spacing, then, as
we see,
the diffusion is very small,  and the tangle can be expected to remain
localized
in the neighbourhood of the grid. If motion on a larger scale is generated,
then the diffusion will be enhanced.

%
% The format of reference should be
% Author1, Author2, Author3, Journal {\bf volume} (year) page.
% No ``and'' between the authors are necessary.
%

\end{document}